\documentclass[]{aa}
\usepackage{graphicx}
\begin{document}
\title{Chemically peculiar stars in the Large Magellanic Cloud\thanks{Based on 
observations at CASLEO and ESO-La Silla}}
\author{E.~Paunzen\inst{1}, H.M.~Maitzen\inst{1}, O.I.~Pintado\inst{2,}\thanks
       {Member of Carrera del Investigador del Consejo Nacional de 
        Investigaciones Cient\'{i}ficas y T\'{e}cnicas de la Rep\'{u}blica 
        Argentina and Visiting Astronomer at Complejo Astron\'omico El Leoncito 
        operated under agreement between the Consejo Nacional de Investigaciones 
        Cient\'{i}ficas y T\'{e}cnicas de la Rep\'ublica Argentina and the 
        National Universities of La Plata, C\'ordoba y San Juan.}, A.~Claret\inst{3}, 
		I.Kh.~Iliev\inst{4}, and M.~Netopil\inst{1}}

\mail{Ernst.Paunzen@univie.ac.at}

\institute{Institut f\"ur Astronomie der Universit\"at Wien,
           T\"urkenschanzstr. 17, A-1180 Wien, Austria
\and	   Departamento de F\'isica, Facultad de Ciencias Exactas 
           y Tecnolog\'ia, Universidad Nacional de Tucum\'an, Argentina - Consejo Nacional 
		   de Investigaciones Cient\'ificas y T\'ecnicas de la Rep\'ublica Argentina		
\and	   Instituto de Astrof\'{\i}sica de Andaluc\'{\i}a
		   CSIC, Apartado 3004, 18080 Granada, Spain   
\and	   Institute of Astronomy, National Astronomical Observatory, 
           P.O. Box 136, BG-4700 Smolyan, Bulgaria
}

\date{Received 31 March 2006 / Accepted 08 July 2006}
\authorrunning{E. Paunzen et al.}{}
\titlerunning{Chemically peculiar stars in the LMC}{}

\abstract{The detection of magnetic chemically peculiar (CP2) stars in open
clusters of extragalactic systems can give observational answers to many
unsolved questions. For example, one can study the influence of different global as well local
environments on the lack of and presence of peculiarities.}
{The mean percentage of CP2 stars in the Milky Way is of the order of 5\% for
the spectral range from early B- to F-type, luminosity class V objects. The origin
of the CP2 phenomenon seems to be closely connected to the overall metallicity and
global magnetic field environment. The theoretical models are still only tested 
by observations in the Milky Way. It is therefore essential to provide high quality
observations in rather different global environments.}
{The young clusters NGC 2136/7 were
observed in the $\Delta a$ photometric system. This intermediate 
band photometric system 
samples the depth of the 520\,nm flux depression by comparing the flux at the center
with the adjacent regions with bandwidths of 11\,nm to 23\,nm.  
The $\Delta a$ photometric system is most suitable for detecting CP2 stars with high
efficiency, but is also capable of detecting a small percentage of 
non-magnetic CP objects. Furthermore, the groups of (metal-weak) $\lambda$ Bootis, 
as well as classical Be/shell stars, can be successfully investigated.}
{We present high precision photometric $\Delta a$ 
observations of 417 objects in NGC 2136/7 and
its surrounding field, of which five turned out to be
bona fide magnetic CP stars. In addition, we discovered two Be/Ae stars.}
{From our investigations of NGC 1711, NGC 1866, NGC 2136/7,
their surroundings, and one independent field of the LMC population, we derive an 
occurrence of classical chemically peculiar stars of 2.2(6)\% 
in the LMC, which is only half the value found in the Milky
Way. The mass and age distribution of the photometrically detected CP stars
is not different from that of similar objects in galactic open clusters.}


\maketitle

\section{Introduction}

The classical chemically peculiar (CP) stars of the upper main sequence 
(luminosity classes V and IV) are targets of detailed investigations since 
their first description by Maury (1897). They provide excellent test 
objects for astrophysical processes like diffusion, convection, and stratification in stellar
atmospheres in the presence of rather strong magnetic fields. These mechanisms
for stars, both in the Milky Way and its surrounding stellar systems, 
contribute important knowledge to stellar evolution under different local circumstances,
e.g., different metallicities. 

CP stars can be detected very efficiently by applying 
the $\Delta a$ photometric system (cf. Paunzen et al. 2005a), which measures the 
characteristic broadband absorption feature located around 520\,nm,
which is most certainly a consequence 
of the non-solar elemental abundance distribution
of CP and related objects in the presence of a strong stellar
magnetic field (Kupka et al. 2004). It 
samples its depth, comparing the flux at the center (521\,nm, $g_{\rm 2}$),
with the adjacent regions (503\,nm, $g_{\rm 1}$ and 551\,nm, $y$), using bandwidths from
11\,nm to 23\,nm. The respective index $a$ was introduced as
$$ a = g_{\rm 2} - (g_{\rm 1} + y)/2. $$
The intrinsic peculiarity index $\Delta a$
is defined as the difference between the individual $a$-values and those 
(=\,$a_{\rm 0}$)
of non-peculiar stars of the same color. The locus of the $a_{\rm 0}$-values
has been called, the normality line.

Virtually all peculiar objects with magnetic fields (CP2 and CP4 stars, Preston 1974) 
have positive $\Delta a$ values in excess of +100\,mmag whereas Be/Ae/shell and 
metal-weak (e.g., $\lambda$ Bootis group) stars exhibit significantly negative ones 
up to $-$35\,mmag (Paunzen et al. 2005a). In general, some spectroscopic binary
systems can mimic a positive $\Delta a$ value of up to +16\,mmag, which is well below
the observed values for the stars presented in this paper. Later type, evolved
objects of luminosity classes III to I might also show the same behavior, but can
be easily identified within the color-magnitude diagram and do not influence
the overall statistics.

In this paper we present our efforts to detect chemically peculiar stars
in the field of NGC 2136/7, a binary cluster system, located in the Large Magellanic
Cloud (LMC). In total, we observed 417 objects of which five turned out to be
bona fide magnetic CP stars. These observations, added to our statistical 
analysis in the LMC presented in Paunzen et al. (2005b), show an overall
occurrence of 2.2(6)\% for chemically peculiar stars in the LMC.

\section{Observations and reduction}

The observations were done on twelve nights at two different
telescopes with the identical $\Delta a$ filter set:
\begin{itemize}
\item CASLEO: 215\,cm telescope, TEK-1024 CCD, field of view
of about 9.5$\arcmin$, August 2001 and January 2003, observer: O.I.~Pintado
\item ESO-LaSilla: Bochum 61\,cm telescope,
Thompson 7882 CCD, 384x576 pixels, 3$\arcmin$\,x\,4$\arcmin$,
April 1995, observers: H.M.~Maitzen and E.~Paunzen
\end{itemize}

The filters have the following characteristics: $g_1$ ($\lambda_c$\,=\,5027\AA,
FWHM\,=\,222\AA, Peak transmission\,=\,66\%),
$g_2$ (5205/107/50), and $y$ (5509/120/54). In total, 69 frames for the three
filters (20/23/26) were obtained.

The bias subtraction, dark correction, flat-fielding, and a point-spread function fitting 
were carried out within standard IRAF V2.12.2 routines on a
Personal Computer with a Linux distribution. 

One of the advantages of the $\Delta a$ photometric system is that
the ``standard'' as well as program stars are always on the same frame.
Because of instrumentally induced offsets and different air masses
between the individual frames, photometric reduction of each frame was 
performed separately and the measurements were then 
averaged and weighted by their individual photometric errors. 

In total, 417 stars were measured in the field. We were not able
to resolve the innermost regions ($\approx$\,20$\arcsec$) 
of NGC2136/7 due to the seeing conditions and saturation.  

The tables with all data for the individual cluster stars as well as
nonmembers are available
in electronic form at the CDS via anonymous ftp to cdsarc.u-strasbg.fr (130.79.128.5),
http://cdsweb.u-strasbg.fr/Abstract.html,
or upon request from the first author. These tables include our internal
numbers, J2000.0 coordinates, the $X$ and $Y$
coordinates within our frames, the observed $(g_{1}-y)$ and
$a$ values with their corresponding errors, the $V$ magnitudes,
the $(B-V)$ colors from the literature, the
$\Delta a$-values derived from the normality lines of $(g_{1}-y)$
(excluding nonmembers), and the number of observations, respectively.

\begin{table}
\caption{Peculiar objects found in the field of NGC2136/7 (upper part)
and its surrounding area (lower part). The objects No. 127 and 228
are most certainly not classical chemically peculiar stars, whereas
No. 219 is a red supergiant with high mass-loss and emission. 
The errors in the final digits of the corresponding quantity are in parenthesis.}
{\scriptsize
\begin{tabular}{rrrlll}
\hline
\hline
\multicolumn{1}{c}{$No$} & \multicolumn{1}{c}{$X$} & \multicolumn{1}{c}{$Y$} & 
\multicolumn{1}{c}{$(B-V)_0$} & \multicolumn{1}{c}{$\Delta a$} & \multicolumn{1}{c}{$M_V$} \\
\hline
165 & +47.79	& $-$87.73	& +0.164(7) &	+0.095 &	$-$0.449(5) \\
204 & +98.00	& $-$16.22	& $-$0.255(8) &	$-$0.062 &	$-$1.480(6) \\
219 & +110.54	& $-$57.90	& +0.479(14) &	$-$0.056 &	$-$4.153(9) \\
228 & +117.09	& $-$30.35	& +0.443(8) &	+0.074 &	$-$4.264(6) \\
233 & +120.22	& $-$214.24	& $-$0.220(2) &	+0.044 & 	$-$1.051(2) \\
283 & +167.19	& $-$96.05	& $-$0.084(3) &	+0.049 &	$-$1.663(2) \\
\hline
91 & $-$91.72   & $-$416.28	& $-$0.205(7) &	+0.041 &	$-$1.483(7) \\
110 & $-$54.52  & $-$157.76	& +0.005(8) &	+0.060 &	+0.051(7) \\
127 & $-$13.32  & $-$146.05	& +0.701(10) &	+0.053 &	+0.330(6) \\
392 & +377.70	& $-$202.34 & +0.033(3) &	$-$0.053 &	$-$0.709(2) \\
\hline
\label{cps}
\end{tabular}
}
\end{table}

\begin{figure*}
\begin{center}
\includegraphics[width=170mm]{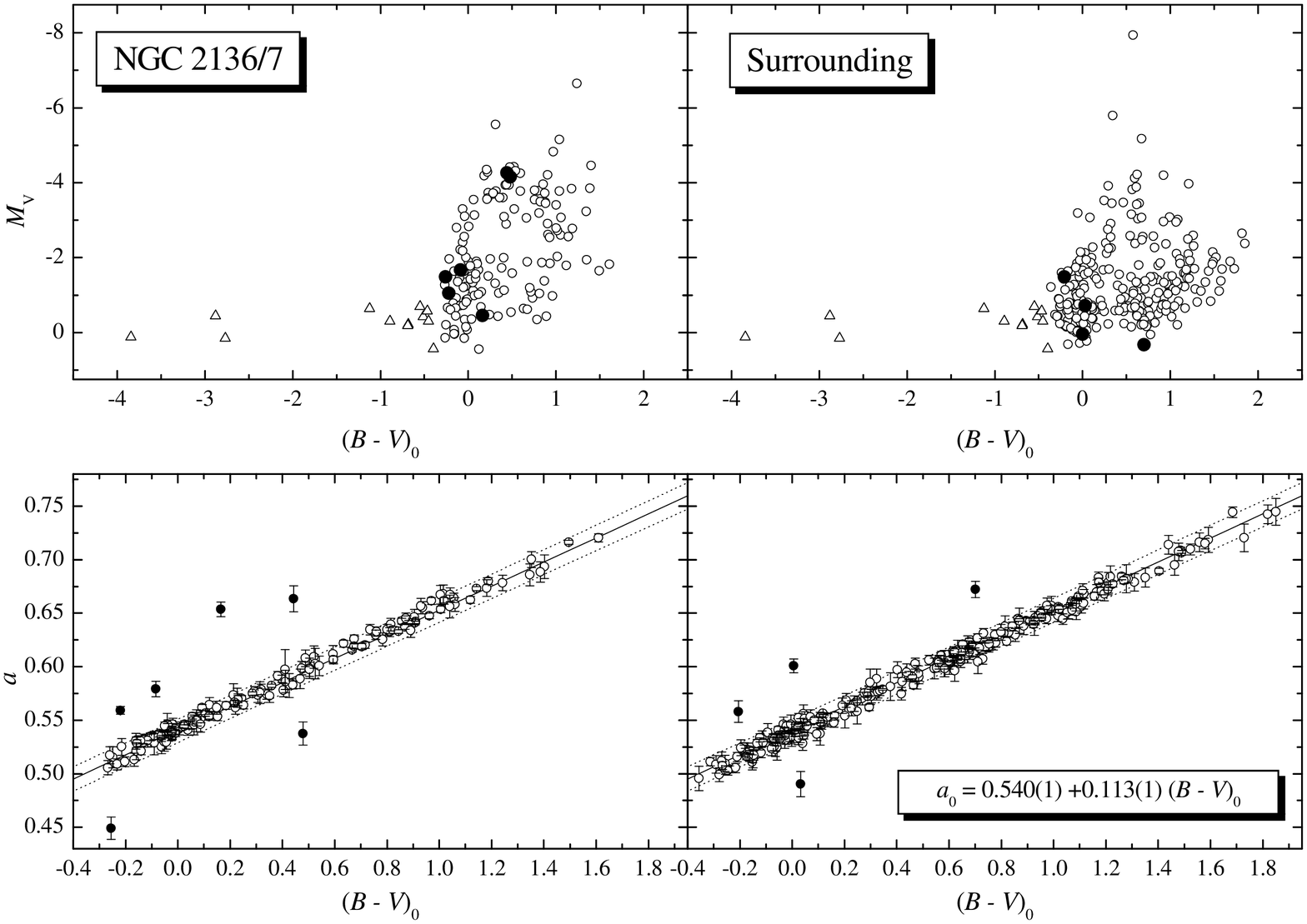}
\caption[]{Observed diagrams for NGC2136/7 (left panels) and
its surrounding area (right panels).
The solid line is the
normality line whereas the dotted lines are the confidence intervals
corresponding to 99.9\%. 
The open circles are members and the filled circles are peculiar stars,
whereas the triangles are non-members.
The error bars for each individual object
are the mean errors. The measurement errors of $M_V$ are much smaller than the
symbols and have been omitted. The $M_V$ and 
$(B-V)_0$ values were calculated using a distance modulus of 18.5\,mag and a reddening
$E(B-V)\,=\,0.1$\,mag.}
\label{fig1}
\end{center}
\end{figure*}

\section{NGC 2136/7 and its surrounding field population}

Dirsch et al. (2000) concluded that NGC 2136 and NGC 2137 are
a physical binary (possibly triple) cluster system of
same age (log\,$t$\,=\,8.0) and metallicity ([Fe/H]\,=\,$-$0.55)
located at 
$\alpha$(2000.0)\,=\,05$^{\rm h}$53$^{\rm m}$00$^{\rm s}$ 
and $\delta$(2000.0)\,=\,$-$69$\degr$29$\arcmin$30$\farcs$
The estimated reddening of $E(B-V)$\,=\,0.1\,mag is typical for
other clusters in the LMC.

Especially interesting is their result of the surrounding field
population. They found a younger age and lower metallicity than
for the clusters. However, the error of their metallicity 
determination is 0.59\,dex.

Unfortunately it was not possible to use the Str{\"o}mgren 
photometry from Dirsch et al. (2000) for the calibration of our 
photometric values and the identification of objects
because it has been
available neither in electronic form nor upon request from the 
authors of the given reference. Furthermore, no printed 
tables are available. 

We have therefore used the Johnson-Kron-Cousins photometric $UBVI$ 
data published by Zaritsky et al. (2004). They provide no consecutive
numbering system but equatorial coordinates (see their Table 1). The
identification of stars in common was done in two steps. First of all,
we identified the brightest objects by eye and derived a first calibration
of the instrumental magnitude $y_{inst}$ versus $V$ and the coordinates within
our frames ($X$, $Y$) versus ($\alpha$, $\delta$). With
these preliminary values, a second automatic iteration was done limiting the
magnitude difference to 0.5\,mag and a varying radius limit. It is difficult
to decide which objects coincide in the cluster areas of NGC 2136/7. If more than
one object from the catalogue by Zaritsky et al. (2004) matches at the same
level of significance, we did not include its $UBVI$ data in the analysis. In total we
identified 248 objects that were used for the final calibration of the
instrumental magnitudes $y_{inst}$ and $(g_1 - y)_{inst}$. In parenthesis 
are the errors in the final digits of the corresponding quantity
\begin{eqnarray}
V &=& -5.372(15) + 0.983(5)\cdot y_{inst} \\
B-V &=& +3.12(6) + 2.38(6)\cdot(g_1 - y)_{inst}.
\end{eqnarray}
The absolute magnitudes were calculated by using the distance modulus of 18.5\,mag taken from 
Alves (2004) and $A_V$\,=\,3.1$\cdot E(B-V)$\,=\,0.31\,mag.

150 apparent members of NGC 2136/7 were selected based on the 
cluster radii and coordinates listed in Dirsch et al. (2000).

For the normality line, the $(g_1-y)$ measurements were converted into
dereddened $(B-V)_0$ values. The normality lines of NGC2136/7 and 
its surrounding area coincide excellently, allowing to merge them into one
final diagram. The following correlation for the normality line
was found to be
\begin{equation}
a_0\,=\,0.540(1)\,+\,0.113(1)\cdot (B-V)_0.
\end{equation}
The result is shown graphically in Fig. \ref{fig1}. The 3$\sigma$ limit is estimated to be
$\pm$0.014\,mag, a value which is well in line with those of galactic open clusters, taking
into account the magnitude limit of about $V$\,=\,19\,mag that was reached 
(Paunzen et al. 2005c).

\section{Results} \label{results}

In Table \ref{cps}, we have summarized the results for the peculiar
objects that deviate significantly from the normality line (Fig.
\ref{fig1}).

From the photometric indices we conclude that the objects No. 91, 110,
165, 233, and 283 fall well into the domain of the classical magnetic chemically 
peculiar objects (G{\'o}mez et al. 1998).

The star No. 228 is a red supergiant with $(B-V)_0$\,=\,+0.443(8) and 
$M_V$\,=\,$-$4.264(6)\,mag. This star group can exhibit 
positive $\Delta a$ values as investigated by Paunzen et al. (2005a).
Also, No. 219 turns out to be a red supergiant, but with the opposite
effect of having a significant negative $\Delta a$ value probably 
caused by a high mass-loss rate together with emission in the stellar envelope 
(van Loon et al. 1999). This might be an effect bearing
some similarity to Be stars (Pavlovski \& Maitzen 1989).

Object No. 127 cannot be addressed as a classical chemically
peculiar star. Its $(B-V)_0$ value of +0.701(10)\,mag and absolute 
magnitude of +0.330(6)\,mag is typical of an evolved G-type star
of luminosity class IV to III (Ginestet et al. 2000). The 
evolutionary status of it is not in line with the age of NGC 2136/7, but it
would be compatible with the ages derived for the field population of the
LMC (Dolphin 2000). There are several peculiar, mainly CN and CH abnormal
stars, in this spectral domain (Taylor 1999). The capability to detect
such objects with $\Delta a$ photometry has not yet been investigated and 
certainly deserves further research.

Finally, No. 204 and 392, which show significant negative $\Delta a$
values, are probable Be/Ae stars because they seem to
be too hot for members of the $\lambda$ Bootis group (Paunzen et al. 2002).

We follow the approach given by Paunzen et al. (2005b) and derive the total
number of chemically peculiar stars having significant positive 
$\Delta a$ values in NGC 2136/7 and its surrounding field
compared to all investigated objects in the relevant spectral range up to F2 or 
$(B-V)_0\,=\,0.3$\,mag resulting in 2.7\% and 1.8\%, respectively. Using these
values and those listed in Table 3 of Paunzen et al. (2005b), we derive a
mean occurrence of 2.2(6)\% for chemically peculiar stars in the LMC. 

North (1993) studied the percentage of chemically peculiar stars in open
clusters of the Milky Way. He found a number of 5\% to 10\% for the magnetic ones 
in open clusters with similar ages to those used for this study (see his Figure III).
This percentage is almost identical with that of the galactic field.
The number for the LMC is only about half the value that in the Milky Way.
Up to now, no correlation of this mean value with the metallicity and the occurrence
in clusters or the field population of the LMC is evident. 

This result is based on four widely different fields in the LMC, making
it highly significant, which is important for theories of the origin and evolution
of CP stars. The overall metallicity of the LMC is reduced by up 
to 0.5\,dex compared to the Sun, and its 
global magnetic field consists of a coherent axisymmetric spiral of field 
strength that is weaker than that of the Milky Way. Both parameters 
might significantly influence the origin of the CP stars' magnetic fields.
Still it is a matter of debate if the stellar magnetic field is due to the survival 
of frozen-in fossil fields originating from the medium out of which the stars were 
formed or if a dynamo mechanism is acting in the stellar interior. 

\begin{figure}
\begin{center}
\includegraphics[width=85mm]{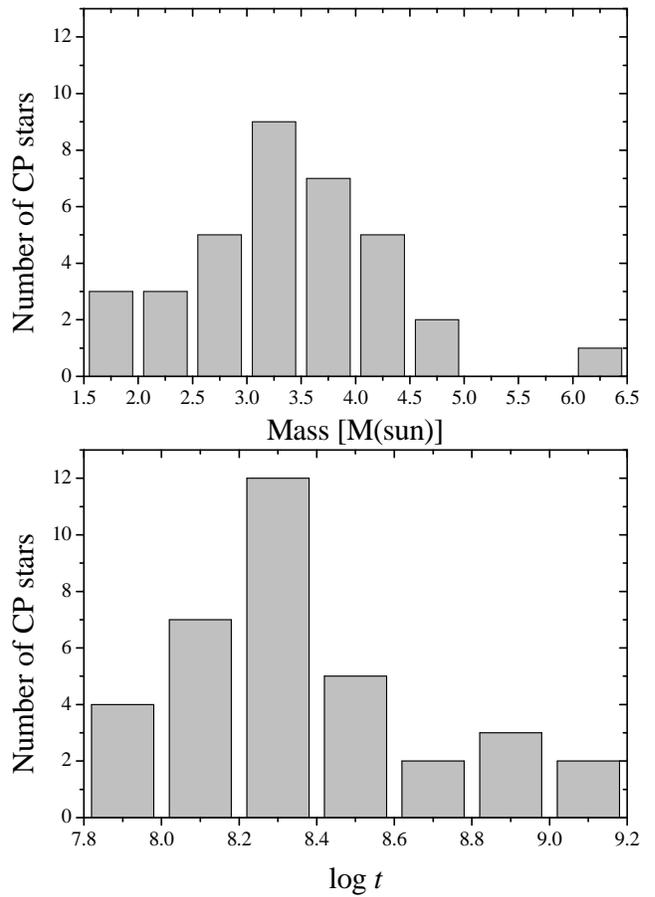}
\caption[]{Mass (upper panel) and age (lower panel) distribution for all
35 photometrically detected CP stars in the LMC. The isochrones for the
analysis are from Claret (2006) who takes the reduced metallicity of the LMC into account.
The distributions are compatible with those of open cluster CP stars in the
Milky Way (North 1993).}
\label{fig2}
\end{center}
\end{figure}

As a last step, we have calibrated the masses and ages for all 35 photometrically
detected CP stars in the LMC using appropriates isochrones by Claret (2006). They take into
account the reduced metallicity and moderate core 
overshooting ([X]\,=\,0.739, [Z]\,=\,0.007, $a_{ov}$\,=\,0.20). All stars were 
individually fitted to derive the age and
mass with a distance modulus of 18.5\,mag as well as a reddening of $E(B-V)\,=\,0.1$\,mag.
Such a ``statistical'' approach naturally introduces an unknown error for the determination
of the masses and ages for the individual objects. But, none of the investigated 
clusters significantly deviates from the chosen values (Maitzen et al. 2001; Paunzen et al.
2005b). From our experience (Claret et al. 2003; P{\"o}hnl et al. 2005), we estimate a heuristic
error of about 15\% for the individual parameters. However, we will only analyze the complete
sample of CP stars and not specific objects. Figure \ref{fig2} shows the histograms of the
ages and masses of the investigated sample of 35 photometrically detected CP stars excluding 
probable cool type objects and Be/Ae as well as metal-weak stars. One has to keep in mind that 
this sample is biased due to poor number statistics and the census of the investigated 
spectral range (= masses). 

The $\Delta a$ values are at a maximum around A0 for CP2 stars and decrease
for hotter and cooler objects, which reflects the incidence of these objects
in respect to all stars in this spectral range (Maitzen \& Vogt 1983). The occurrence of hot
CP4 stars compensates, to a certain amount, the overall percentage for early to late
B-type objects (Paunzen et al. 2005a). This behavior can be also seen in Fig. \ref{fig2}.

If we compare Fig. \ref{fig2} to the results for CP stars in galactic
open clusters by North (1993), we find perfect agreement for the following conclusion: {\it
all kinds of CP stars are main sequence, core-hydrogen burning objects with masses between 1.5
and 7\,M$_{\sun}$.}

\section{Conclusions}

We present the results of our search for CP stars of the upper main sequence in the LMC 
clusters NGC2136/7 and the surrounding field applying the 
intermediate band $\Delta a$ photometry, which measures the 
characteristic broadband absorption feature located around 520\,nm. 

In total, 417 objects were measured on 69 individual frames observed
with two different telescopes. We report the detection of five classical chemically
peculiar objects and two Be/Ae stars. In addition, three peculiar objects were
found that deserve further attention.

We conclude from our investigations of NGC 1711, NGC 1866, NGC 2136/7,
their surroundings, and one independent field of the LMC population that the 
occurrence of classical chemically peculiar stars is 2.2(6)\% for chemically 
peculiar stars in the LMC.

The age and mass distributions, derived by applying appropriate
isochrones, do not alter from those of CP stars
in galactic open clusters.

This provides a valuable observational source for understanding the CP phenomenon
of the upper main sequence in a different global environment than in our Milky Way.
  
\begin{acknowledgements}
This research was performed within the projects  
{\sl P17580} and {\sl P17920} of the Austrian Fonds zur F{\"o}rderung der 
wissen\-schaft\-lichen Forschung (FwF), as
well as the City of Vienna (Hochschuljubil{\"a}umsstiftung project: $\Delta a$ 
Photometrie in der 
Milchstrasse und den Magellanschen Wolken, H-1123/2002). 
I.Kh.~Iliev acknowledges support by the Bulgarian National Science Fund under grant F-1403/2004
and the Pinehill Foundation.
Use was made of the SIMBAD database, operated at the CDS, Strasbourg, France and
the WEBDA database, operated at the University
of Vienna. This research has made use of NASA's Astrophysics Data System.
The authors acknowledge the use of the CCD and data reduction acquisition system 
supported by US NSF Grant AST 90-15827 to R.M.~Rich.
\end{acknowledgements}

\end{document}